\documentclass[runningheads,a4paper]{llncs}

\usepackage{cmap}
\usepackage{latexsym}
\usepackage[cp1251]{inputenc}
\usepackage[unicode]{hyperref}

\usepackage{amssymb}
\usepackage{graphicx}

\usepackage{tikz}
\usetikzlibrary{shapes,arrows, positioning}
\pgfmathtruncatemacro\distance{1}

\usepackage{url}
\urldef{\mailsa}\path|amironov66@gmail.com|
\urldef{\mailsb}\path|anna.kramer, leonie.kunz, christine.reiss, nicole.sator,|
\urldef{\mailsc}\path|erika.siebert-cole, peter.strasser, lncs}@springer.com|

\begin{document}
\def\enemy{{\dagger}}
\def\xor{
\oplus
}

\def\eqs#1{\mathop{\sim}\limits_{#1}}

\def\event#1{#1}

\def\simplus{
\approx
}

\def\ite#1#2#3{\left\{\begin{array}{rlll}
[\![{#1}]\!]&{#2}\\
:&{#3}
\ey\right.}

\def\g#1{\left[\!\!\left[
{\def\arraystretch{1}
\begin{array}{lllll}#1\end{array}}
\right]\!\!\right]}

\def\bcy{\begin{array}{ccc}}
\def\pdownl#1{
  \begin{picture}(8,20)
  \put (4,20){\vector(0,-1){30}}
  \put (1,6){\makebox(1,1)[r]{$\scriptstyle #1$}}
  \end{picture} }
  \def\pleft#1{
  \begin{picture}(20,8)
  \put (25,3){\vector(-1,0){30}}
  \put (12,10){\makebox(1,1){$\scriptstyle #1$}}
  \end{picture} }
  \def\pdownr#1{
  \begin{picture}(8,20)
  \put (4,20){\vector(0,-1){30}}
  \put (7,6){\makebox(1,1)[l]{$\scriptstyle #1$}}
  \end{picture} }

\def\dey#1#2{#1 (#2)}
\def\deyc#1#2{#1 \cdot  #2}
\def\bcy{\begin{array}{ccc}}

\def\ral#1{\;\mathop{\longrightarrow}\limits^{#1}\;}
\def\bc{\begin{center}\begin{tabular}{l}}
\def\ec{\end{tabular}\end{center}}
\def\modn#1{\mathop{=}\limits_{#1}}
\def\inn#1{\mathop{\in}\limits_{#1}}
\def\modnop#1{\mathop{#1}\limits_{n}}
\def\rar{\mathop{\in}\limits_{r}}
\def\und#1{\mathop{=}\limits_{#1}}
\def\skobq#1{\langle\!| #1 |\!\rangle}
\def\redeq{\;\mathop{\approx}\limits^{r}\;}
\def\reduc{\;\mathop{\mapsto}\limits^{r}\;}
\def\pt{\;\mathop{+}\limits_{\tau}\;}
\def\sost{\begin{picture}(0,0)\put(0,3){\circle*{4}}
\end{picture}}
\def\sosto{\begin{picture}(0,0)\put(0,0){\circle*{4}}
\end{picture}}
\def\bi{\begin{itemize}}
\def\pa{\,|\,}
\def\oc{\;\mathop{\approx}\limits^{+}\;}
\def\p#1#2{(\;#1\;,\;#2\;)}
\def\mor#1#2#3{\by #1&\pright{#2}&#3\ey}
\def\ei{\end{itemize}}
\def\bn{\begin{enumerate}}
\def\en{\end{enumerate}}
\def\i{\item}
\def\a{\forall\;}

\def\l#1{[#1]}
\def\ll#1{[\![#1]\!]}
\def\lc#1{\langle#1\rangle}
\def\lcc#1#2{{#1}^{#2}}

\def\fm#1{\left[\!\!\left[\begin{array}{lllll}#1\end{array}\right]\!\!\right]}
\def\ra#1{\mathop{\to}\limits^{\!\!#1}}
\def\dra#1{\mathop{\Rightarrow}\limits^{\!\!\!\!#1}}

\def\bigset#1#2{\left\{\by #1 \left| \by #2 \ey\right\}\ey\right.}
\def\p{\leftarrow}

\def\pmiddleright#1{
  \begin{picture}(30,18)
  \put (-5,3){\vector(1,0){40}}
  \put (12,8){\makebox(1,1){$\scriptstyle #1$}}
  \end{picture} }

\def\plongright#1{
  \begin{picture}(40,18)
  \put (-5,3){\vector(1,0){50}}
  \put (20,8){\makebox(1,1){$\scriptstyle #1$}}
  \end{picture} }

\def\plongleft#1{
  \begin{picture}(40,8)
  \put (45,3){\vector(-1,0){50}}
  \put (20,8){\makebox(1,1){$\scriptstyle #1$}}
  \end{picture} }

\def\pse#1#2{
  \begin{picture}(40,8)
  \put (-5,-5){\vector(2,-1){50}}
  \put (45,-5){\vector(-2,-1){50}}
  \put (-5,-12){\makebox(1,1)[r]{$\scriptstyle #1$}}
  \put (45,-12){\makebox(1,1)[l]{$\scriptstyle #2$}}
  \end{picture} }

\def\und#1{\mathop{=}\limits_{#1}}
\def\redeq{\;\mathop{\approx}\limits^{r}\;}
\def\reduc{\;\mathop{\mapsto}\limits^{r}\;}
\def\oc{\mathop{\approx}\limits^{+}}
\def\sost{\begin{picture}(0,0)\put(0,0){\circle*{4}}
\end{picture}}
\def\bi{\begin{itemize}}
\def\pa{\,|\,}
\def\oo{\;\mathop{\approx}\limits^{c}\;}
\def\p#1#2{(\;#1\;,\;#2\;)}
\def\mor#1#2#3{\by #1&\pright{#2}&#3\ey}
\def\ei{\end{itemize}}
\def\bn{\begin{enumerate}}
\def\en{\end{enumerate}}
\def\i{\item}
\def\bigset#1#2{\left\{\by #1 \left| \by #2 \ey\right\}\ey\right.}
\def\p{\leftarrow}
\def\buffer{{\it Buffer}}
\def\eam{\mathbin{{\mathop{=}\limits^{\mbox{\scriptsize def}}}}}
\def\be#1{\begin{equation}\label{#1}}
\def\ee{\end{equation}}
\def\re#1{(\ref{#1})}

\def\bn{\begin{enumerate}}
\def\en{\end{enumerate}}
\def\bi{\begin{itemize}}
\def\ei{\end{itemize}}
\def\i{\item}
\def\c#1{\left[\!\!\!\!\!\by\left[
{\def\arraystretch{1}
\begin{array}{lllll}#1\end{array}}\right]\ey\!\!\!\!\!\right]}
\def\d#1{\left[\begin{array}{lllll}#1\end{array}\right]}
\def\b#1{\left(\begin{array}{lllll}#1\end{array}\right)}
\def\ra#1{\;\mathop{\to}\limits^{\!\!#1}\;}
\def\leqd{\;\mathop{<}\limits_{2}\;}
\def\diagrw#1{{
  \def\normalbaselines{\baselineskip20pt \lineskip3pt \lineskiplimit3pt }
  \matrix{#1}}}

\def\blackbox{\vrule height 7pt width 7pt depth 0pt}
\def\pu#1#2{
\mbox{$\!\!\begin{picture}(0,0)
\put (-#1,-#2){\line(1,0){#1}}
\put (-#1,-#2){\line(0,1){#2}}
\put (#1,#2){\line(-1,0){#1}}
\put (#1,#2){\line(0,-1){#2}}
\put (-#1,#2){\line(1,0){#1}}
\put (-#1,#2){\line(0,-1){#2}}
\put (#1,-#2){\line(-1,0){#1}}
\put (#1,-#2){\line(0,1){#2}}
\end{picture}$}
}

\def\pright#1{
  \begin{picture}(20,18)
  \put (-5,3){\vector(1,0){30}}
  \put (9,10){\makebox(1,1){$\scriptstyle #1$}}
  \end{picture} }

\def\by{\begin{array}{llllllllllllll}}
\def\ey{\end{array}}

\setcounter{theorem}{1}


\title{Verification of MPI programs}

\author{Andrew M. Mironov
}

\date{}

\institute{Moscow State University,
Faculty of Mechanics and Mathematics\\
\mailsa
}

\maketitle

\begin{abstract}
In this paper, we outline an approach to verifying parallel programs. 
A new mathematical model of parallel programs is introduced.
The introduced model is illustrated by the verification of the  matrix multiplication MPI program. 

\end{abstract}

\section{Introduction}

{\bf Parallel programs} are computer programs designed to 
be executed on multiprocessor computing systems (MPCS).
The problem of developing correct and safe parallel programs is currently 
highly  topical. Formal verification of  correctness and safety properties of parallel programs is a complex mathematical problem. The existing methods for solving this problem are suitable only for a  limited class of parallel programs.

One of the most widely used languages for describing parallel programs is MPI (Message Passing Interface).

In this paper, a new mathematical model of MPI programs is introduced.
On the basis of this model one can solve the problems of verifying parallel programs presented on a certain subset of MPI.
The introduced model is illustrated by the verification of the  matrix multiplication MPI program. 

The most essential feature of the approach to modeling and verification of MPI programs presented in this paper
is the possibility of using this approach for MPI 
programs that can generate any number of processes. 

Among other approaches to modeling and verifying MPI programs for any number of processes it should be noted the approach 
in the work  \cite{protbased}.
In this work the tool for modeling and verifying MPI programs  called ParTypes is presented.
It requires the user to provide a protocol which specifies the communication pattern of an execution.  Unfortunately, it has some limitations.  In particular, it does not work for wildcard receives, which means it cannot be applied to the 
MPI program for 
matrix multiplication considered in the present paper.
There are other approaches using symbolic execution and model checking 
(\cite{1}-\cite{8}), but all of them require a bound on the number of processes.

\section{Essentials for MPI }

\subsection{MPI programs}

{\bf MPI (Message Passing Interface)} is a set of functions, types and constants for a development of parallel programs (called {\bf MPI pro\-g\-rams}). A MPI pro\-g\-ram is a C program in which functions, types and constants from MPI can be used.
An
execution of an MPI pro\-g\-ram on a MPCS has the following form:
at each node of the MPCS, a computational process corresponding to this MPI pro\-g\-ram is generated. All processes generated by the MPI program operate in parallel and can exchange information with each other through {\bf message passing}. 

Each process generated by a MPI program has a {\bf rank}, 
which is a number from the set $ \{0, \ldots, m-1 \} $, where $ m $ is the number of processes generated by the  MPI program. A process with rank 0 is called a {\bf root process}.

MPI has  functions \verb'MPI_Comm_rank 'and \verb'MPI_Comm_size'.
Each process generated by a MPI program can use these functions 
to find out its rank and a number of  processes generated by this MPI program, respectively: \bi\i
\verb'MPI_Comm_rank (MPI_COMM_WORLD, &rank);'\\
after executing this function, a value of variable \verb'rank' (\verb'int') 
will be equal to the rank of the process that called this function,
\i
\verb'MPI_Comm_size (MPI_COMM_WORLD, &nprocs);'\\
after executing this function, a value of variable \verb'nprocs' (\verb'int')
will be equal to the number of processes generated by the MPI program. \ei

\subsection{Message Passing   Functions }
\label{peredachasoob}

In MPI, a {\bf message} is an array of data of a certain type, and {\bf message passing (MP)} is an action, as a result of which a message is sent by one process  and is received by 
 other process (or processes). 
A sent message is placed in a queue, from which it will then be taken by the receiving process.
 
We will consider the following types of
MP MPI functions: \bi \i
 {\bf pairwise MP (PMP)}:
there are two processes involved in a PMP: a sender of a message, and 
 a receiver of this message,
 \i {\bf broadcast MP (BMP)}:
all processes generated by a MPI program participate in 
 an executing of a BMP,
 root process is a sender of a message, and other processes 
are  receivers of this message.
\ei

Messages sent by PMP functions cannot be received by BMP functions, and vice versa.
Below we describe some of  MP  functions.
In the descriptions, for each argument of these functions, we indicate its type in parentheses. 

\bn
\i Sending a message (PMP) :
\be{sdfsfsdfsend}
\by
{\tt MPI\_Send}& (p, n, \tau,
r, l, {\tt MPI\_COMM\_WORLD});\ey\ee
This function is performed by sending a message to a process with rank $ r $ (\verb'int'). The message being sent is an array of $ n $ (\verb'int') elements of type $ \tau $ (\verb'MPI_Datatype'), the beginning of which is at $ p $ (\verb'void *'). The tag (i.e. label) $ l $ (\verb'int') is appended to the message being sent.
\i Receiving a message (PMP): 
\be{sdfsfsdfreceive}
\by{\tt MPI\_Recv}&(p, n, \tau,{\tt MPI\_ANY\_SOURCE}, 
{\tt MPI\_ANY\_TAG}, {\tt MPI\_COMM\_WORLD}, q);\ey\ee

This function is performed by receiving a message, which should be placed in the memory location the beginning of which is at 
$ p $. It is assumed that the received message is an array of no more than $ n $ elements of type $ \tau $.
 
$ q $ \verb'(MPI_Status *)' is an address of a structure in which information about the received message should be placed. This structure contains the following fields: \verb'MPI_SOURCE' (the sender's rank must be placed in it), \verb'MPI_TAG' (the received message tag must be placed in it), and other fields.
\i Sending a message from a root process to other processes (BMP): 
\be{sdfsfsdfbcast}
\by{\tt MPI\_Bcast}& (p, n, \tau, 0, {\tt MPI\_COMM\_WORLD});\ey\ee
This function is performed as follows.
\bi
\i In a root process, a message is sent to all other processes, which is an array of $ n $ (\verb'int') elements of type $ \tau $ (\verb'MPI_Datatype'), the beginning of which is located at $ p $ (\verb'void *').
\i In other processes, a message is received from the root process, which 
must be located in the memory location at $ p $. It is assumed that the received message is an array of no more than $ n $ elements of type $ \tau $.
\ei
\en

\section{ Matrix Multiplication MPI Program }
\label{rejgkljktdhgdslfk}

In this section, we present an example of a matrix multiplication MPI program. 
The example is taken from 
\cite{matrixmult}. 
This example is used below to illustrate the application of the model of MPI programs described in this work for verifying MPI programs. 

\subsection{Informal Description of Matrix Multiplication MPI Program}

The problem of matrix multiplication is to calculate a product $ C = AB $ given the matrices $ A $ and $ B $. 
Informally, the work of the MPI pro\-g\-ram $\Pi$
for multiplying the matrices $ A $ and $ B $, stated in this section, can be described as follows. We will call a root process of $\Pi$ a {\bf manager} and  other  processes of $\Pi$ {\bf workers}. Manager's job consists of the following actions:
\bi
\i sending  second matrix ($ B $) to all workers,
\i assigning tasks to workers, and receiving results from workers.
\ei

Each task for a worker is to calculate one row of the matrix $ C = AB $. The manager assigns this task by sending the worker a message containing one row of matrix $ A $. A tag of this message is equal to the number of the row being sent.
As soon as the manager receives a result from a worker (i.e., a message with the calculated row of the product, its tag is equal to the number of this row), he sends this worker
either a new task (if there are still unassigned tasks), or a message with tag 0 (if there are no unassigned tasks). 

 \subsection{MPI Matrix Multiplication Program}
 \label{sdfdsfg3wergtegr}

The following  matrix multiplication MPI program $\Pi$ 
uses  auxiliary function \verb'vecmat' to multiply the row \verb'vector[L]' by the matrix 
\verb'matrix[L][M]' and write the result to the array \verb'result[M]'. This function looks like this: 
\begin{verbatim}
void vecmat (double vector[L],
             double matrix[L][M],
             double result[M])
{ int j, k;
  for (j = 0; j < M; j++)
    for (k = 0, result[j] = 0.0; k < L; k++)
      result[j] += vector[k]*matrix[k][j];
}
\end{verbatim}

In the MPI program $\Pi$ presented below, we  use the following notation: \verb'input(a, b);' and \verb'output(c);' are abbreviations of the functions for reading from the file of the factors and writing to the file of the product, respectively.

A MPI program $\Pi$  for multiplying matrices $ A $ and $ B $ has the following form (in this program, to the left of each line, we indicate its number, this is necessary to describe the correspondence between  components of this program and  components of the model ${\cal P}_\Pi$ of this program): 
\begin{verbatim}
01 #define comm MPI_COMM_WORLD
02
03 int main(int argc, char *argv[])
04 { int rank, nprocs, i, j;
05   MPI_Status status;
06 
07   MPI_Init(&argc, &argv);
08   MPI_Comm_size(comm, &nprocs);
09   MPI_Comm_rank(comm, &rank);
10   
11   if (rank == 0)
12   { int count;        
13     double a[N][L], b[L][M], c[N][M], tmp[M];
14 
15     input(a, b);
16     MPI_Bcast(b, L*M, MPI_DOUBLE,
17               0, comm);
18     for (count = 0;
19          count < nprocs-1 && count < N;
20          count++)
21       MPI_Send(&a[count][0], L, MPI_DOUBLE,
22                count+1, count+1, comm);
23     for (i = 0; i < N; i++)
24     { MPI_Recv(tmp, M, MPI_DOUBLE,
25                MPI_ANY_SOURCE, MPI_ANY_TAG,
26                comm, &status );
27       for (j = 0; j < M; j++)
28         c[status.MPI_TAG-1][j] = tmp[j];
29       if (count < N)
30       { MPI_Send(&a[count][0], L, MPI_DOUBLE,
31                  status.MPI_SOURCE, count+1, comm);
32         count++;
33       }
34     }
35     for (i = 1; i < nprocs; i++)
36       MPI_Send(NULL, 0, MPI_INT,
37                i, 0, comm);
38     output (c);
39   }
40   
41   else
42   { double b[L][M], in[L], out[M];
43   
44     MPI_Bcast(b, L*M, MPI_DOUBLE,
45               0, comm);
46     while (1)
47     { MPI_Recv(in, L, MPI_DOUBLE,
48                0, MPI_ANY_TAG,
49                comm, &status);
50       if (status.MPI_TAG == 0) break;
51       vecmat(in, b, out);
52       MPI_Send(out, M, MPI_DOUBLE,
53                0, status.MPI_TAG, comm);
54     }
55   }
56 
57   MPI_Finalize();
58   return 0;
59 }
 \end{verbatim}
 
 \subsection{Auxiliary notation}
\label{sadfgsrtrh465e}
 
For the convenience of modeling and verification this program, we  introduce special designations for some of the objects used in it:
\bi
\i arrays \verb'a[N][L]',
\verb'b[L][M]',
\verb'c[N][M]',  will be denoted by symbols $ A, B, C $ and interpreted as corresponding matrices,
\i the message sent by the \verb'MPI_Send' functions in lines 21, 22 and 30, 31 of the program will be understood as the corresponding row of the matrix $ A $, and denoted by  $ A_i $, where $ i = {{\tt count} +1} $, \i array \verb'c[status.MPI_TAG-1][M]', into which the message received by the \verb'MPI_Recv' function in lines 24, 25, 26 is copied, we  interpret  it as a corresponding row of $ C $, and denote it by  $ C_l $, where $ l = $ \verb'status.MPI_TAG',
 \i from the definition of the \verb'vecmat' function it follows that 
 array  \verb'out' calculated as a result of the execution of the \verb'vecmat'
 function on line 51 corresponds to the product of the row $ Y $ corresponding to array  \verb'in' by matrix $ B $, we will denote this array \verb'out' in the model of $\Pi$ by the product  $ YB $.
 \ei 
  
\subsection{A specification of the  matrix multiplication MPI program}\label{sdafadsfgaetga}
 
A specification of the   MPI program $\Pi$
is the following statement:
after a completion of any execution of $\Pi$ the equality 
$ C = AB $ holds, i.e.
\be{sdfvdsgg3r435grasdv}
\forall\,i=1,\ldots, N\quad
C_i=A_iB.\ee

\section{A model of a MPI program}

In this section, we introduce a concept of a model of 
a  MPI program. This model is designed for formal  
representation of MPI programs that use the above message passing functions.
The basic concepts of this model are sequential and distributed processes. A sequential process is a model of a computational process generated by a MPI program on a node of a MPCS, and a distributed process is a model of a  MPI program on the whole.
The proposed model is a theoretical basis for solving problems of verifying MPI programs.


\subsection{Auxiliary concepts }

We assume that there are given 
 sets $ {\cal T} $, $ {\it {\it {\cal X}}} $ and $ {\cal F} $,  elements of which are called {\bf types}, {\bf variables}, and {\bf function symbols (FS)}, respectively.
Each element $ x $ of $ {\it {\cal X}} $ and $ {\cal F} $ is associated with some type $ \tau_x \in {\cal T} $. $ \forall \, f \in {\cal F} $ $ \tau_f $ has the form 

\be{sdfsdfsfsafe3q}
(\tau_1,\ldots,\tau_n)\to \tau,\quad\mbox{where }
\tau_1,\ldots, \tau_n, \tau \in {\cal T}.\ee

$\forall\,\tau\in {\cal T}$
a set $ {\cal D} _ \tau $ of {\bf values} of type $ \tau $ is given.  
 $ {\cal D} $ denotes a set of values of all types.
$ {\cal T} $ has the  types {\bf B}, {\bf N}, {\bf C}, where
$ {\cal D} _ {\bf B} = \{0,1 \} $,
$ {\cal D} _ {\bf N}$ is the set of natural numbers $\{0,1, \ldots \} $,
 values of type {\bf C} are called {\bf channels}.

$ \forall \, f \in {\cal F} $, if $ \tau_f $ has  form \re {sdfsdfsfsafe3q}, then this FS is associated with a function (denoted by the same symbol $ f $) of the form
 $
{\cal D} _ {\tau_1} \times \ldots \times {\cal D} _ {\tau_n} \to
 {\cal D} _ {\tau} $.

The set $ {\cal E} $ of {\bf terms} is defined inductively.
Each term $ e \in {\cal E}$ is associated with a type $ \tau_e \in {\cal T} $.
The definition of a term is as follows:
each $e\in {\cal D}\cup{\cal X}$ is a term of the type $\tau_e$,
and if $ f \in {\cal F} $, $ e_1, \ldots, e_n \in {\cal E}$, and $ \tau_f = (\tau_ {e_1}, \ldots, \tau_ {e_n}) \to \tau, $ then $ f (e_1, \ldots, e_n) $ is a term of type $ \tau $.

$\forall\,e\in {\cal E}\;\; {\cal X} _e= \{x \in {\cal X} \mid
     x \mbox { occurrs in } e \}$,
${\cal E}_0=\{e\in {\cal E}\mid{\cal X} _e = \emptyset\}$.
Each 
$e \in {\cal E}_0 $ is associated with a {\bf value} $ value (e) \in {\cal D}$, where
 $\forall\, e \in {\cal D}\;\; value (e) = e $, and
 if $ e = f (e_1, \ldots, e_n) \in {\cal E}_0$, then $ value (e) = f (value (e_1), \ldots, value (e_n)) $. 
$\forall\,e\in {\cal E}_0$ the value $value(e)$ will be denoted by the
same notation $e$.

We will assume that \bi \i $ \forall \, n \geq 1 $ $ {\cal F} $ has  FS $ tuple_n $, which allows to construct tuples: for each list of terms $ e_1, \ldots, e_n $, the set $ { \cal E} $ has the term $ tuple_n (e_1, \ldots, e_n)$, which we will denote by $ (e_1, \ldots, e_n) $ and interpret as a tuple of terms $ e_1, \ldots, e_n $,
\i $ {\cal F} $ has  FS $ channel $ of the type $ {\bf N} \to {\bf C} $, $ \forall \, i \geq 0 $ the channel $ channel (i) $ is said to be an $ i $-th channel, a term of the form $ channel (e) $ will be denoted by  $ c_e $,
\i $ {\cal D} _ {\bf C} $ has a {\bf broadcast channel}
$ \circ $, it differs from all channels $ c_i \;(i\geq 0)$.
\ei

$ \forall \, E \subseteq {\cal E} \; \;
 \forall \, \tau \in {\cal T}
\; \; E_ \tau = \{e \in E \mid \tau_e = \tau \} $.
The sets $ {\cal E} _ {\bf B} $ and $ {\cal E} _ {\bf C} $
are denoted by $ {\cal B} $ and $ {\cal C} $, elements of 
${\cal B}$ are called {\bf formulas}.
 $ \forall \, X \subseteq {\it {\cal X}} \; \;
{\cal E} (X) = \{e \in {\cal E} \mid
{\it {\cal X}} _ e \subseteq X \} $,
$ {\cal B} (X) = {\cal E} (X) \cap {\cal B} $.
Below, for each function $ f: E \to E'$ under consideration, where $ E, E' \subseteq {\cal E} $, we  assume that $ \forall \, e \in E \; \; \tau_ {f (e)} = \tau_e $.

$ {\cal D} ^ * $ denotes the set of all tuples of the form $ (d_1, \ldots, d_n) $, where $ n \geq 0 $ and $ d_1, \ldots, d_n \in {\cal D} $, if $ n = 0 $ 
then 
the corresponding tuple is said to be {\bf empty} and is
denoted by $ \varepsilon $.
Elements of $ {\cal D} ^ * $ are called {\bf queues}. 
$ \forall \, D\in {\cal D} ^ * $ 
$ | D | $ denotes the number of components in $ D $. 
If $ D \in {\cal D} ^ * $ and $ 1 \leq i \leq |D| $, 
then $ D_i $ denotes  $ i $-th component of $D$. 
$ \forall \, M \subseteq {\cal D} ^ * $, $ \forall \, i \geq 1 \;\;
 M_ {i} $ denotes the set of $ i $-th components 
of tuples from $ M $.
If $ D = (d_1, \ldots, d_n) \in {\cal D} ^ * \setminus\{\varepsilon\}$, 
then  $ head (D) $ and $ tail (D) $ denote the value 
$ d_1$ and the queue $ (d_2, \ldots, d_n) $ respectively. 

A {\bf binding} is a function $ \theta: {\it {\cal X}} \to {\cal E} $.
The set of all bindings is denoted by $\Theta$.

We will use the following notation: 
\bi 
\i $ \forall \, X \subseteq {\it {\cal X}} \quad \Theta (X) =
\{\theta \in \Theta \mid \forall \, x \in {\it {\cal X}} \setminus X \; \; \theta (x) = x \}, $
\i
 $ \forall \, \theta \in \Theta, \; \forall \,
e \in {\cal E} \;\; e ^ {\theta} $ is a term obtained from $ e $ by replacing $ \forall \, x \in {\it {\cal X}} _ e $ each occurrence of $ x $ in $ e $ on the term $ \theta (x) $,
\i
$ \forall \, \theta, \theta'\in \Theta \;\;
\theta \theta' $ is a binding such that
$ \forall \, x \in {\cal X} \; \; (\theta \theta ') (x) = (x ^ {\theta}) ^ {\theta'} $.
\ei

\subsection {Sequential processes}
\label{sdfdsfgdsgsersdvzx}

In this section we define a concept of a sequential process (SP). A SP is a model of a computational process generated by a MPI program on a node of a MPCS.

{\bf Elementary actions (EA)} are notations of the following forms:
$$ \by
c! e, \quad c? e, \quad e: = e ', \quad
[\! [\varphi] \!], \quad
\mbox {where }\; c \in {\cal C}, \; e, e '\in {\cal E},
\tau_e = \tau_ {e '},
\varphi \in {\cal B}, \ey $$
which are called a {\bf sending} message $ e $ to channel $ c $, a {\bf receiving} message $ e $ from channel $ c $, an {\bf assignment}, and 
a {\bf conditional transition}, respectively.

An
{\bf action} is a finite sequence of EAs, in which there is no more than one sending or receiving. An action $\alpha$ is called a {\bf sending} or a {\bf receiving} if one of  EAs occurred in $\alpha$ is a sending or a receiving, respectively. An action that is not a sending or a receiving is called an {\bf internal action}. Each EA can be considered as an action consisting of this EA.
The set of all actions is denoted by $ {\cal A} $.
$ \forall \, \alpha \in {{\cal A}} \;\; {\it {\cal X}} _ {\alpha}$ is
 the set of all variables occurred in $ \alpha $.
$\forall\, \theta \in \Theta, \forall\, \alpha \in {\cal A}\;\;
 \alpha ^ {\theta} $ denotes an action obtained from $ \alpha $ by replacing each $ x\in {\cal X}_\alpha $ on $ x ^ \theta $. 
 
A {\bf sequential process (SP)} is a triple $ (P, X, \varphi) $,  components of which have the following meaning:
\bi \i $ P $ is a graph with a selected node $ P ^ 0 $ (called an {\bf initial node}), each edge of which has a label $ \alpha \in {\cal A} $,
\i $ X \subseteq {\it {\cal X}} $ is a set {\bf of input variables} of SP $ P $, and
\i $ \varphi \in {\cal B} $ is an {\bf initial condition} of SP $ P $.
\ei

For each SP $ (P, X, \varphi) $
\bi
\i this SP is denoted by the same symbol $ P $ as a graph of this SP, 
the set of nodes of the graph $ P $ is also denoted by $ P $,
\i $ X_P $ and $ \varphi_P $ denote the second and third components of $ P $ respectively, \i $ {\it {\cal X}} _ P $ is the set of all variables occurred 
in $ P $, \i
  $ \hat X_P $ denotes the set $ {\cal X} _P \setminus X_P $ of
  {\bf private} variables of SP $P$,
  \i $ {\cal A} _P $ denotes the set of labels of edges of $ P $,
  \i
$ P ^ {v \to v '} $ denotes an  edge of  $ P $ from $ v $ to $ v' $.
\i each node of the graph $ P $ is an element of the set $ {\cal D} $,
\i $ P $ contains private variable $ at_P $, and for each edge of the graph $ P $, if $ v $ and $ v '$ are the start and the end of this edge, respectively, then the first EA in the label of this edge has the form $ [ \! [at_P = v] \!] $, and the last one is $ at_P: = v '$, these EAs will not be specified explicitly.
\ei 


A SP is a formal description of a behavior of a dynamic system, a work of which is a sequential execution of actions.

A {\bf state} of SP $ P $ is a pair
$s = (
\theta ^ s,
\{[c] ^ s \mid c \in {\cal D} _ {\bf C} \})$,
where \bi
\i $ \theta ^ s \in \Theta ({{\cal X} _P}) $ is a binding, 
such that $ \forall \, x \in {\cal X} _P \; \; {\theta ^ s}(x)\in {\cal E}_0 $,
\i $ \forall \, c \in {\cal D} _ {\bf C} \; \; [c] ^ s \in {\cal D} ^ * $ is
a queue called a {\bf content} of channel $ c $ in state $ s $.
\ei

The set of all states of  SP $ P $ is denoted by $ \Sigma_P $.

$\forall\, s \in \Sigma_P $, $\forall\,  e \in {\cal E} ({\cal X} _P) $, the value $ e ^ {\theta ^ s} $ is denoted by $ e ^ s $.

 A state $s\in \Sigma_P $  is said to be 
{\bf initial}, if 
$s= (
\theta, \{\varepsilon \mid c \in {\cal D} _ {\bf C} \}) $,
where $ \varphi_P ^ {\theta} = 1 $ and $ at_ {P} ^ s = P ^ 0 $.
An initial state of $P$ is denoted by $ 0_P $.
A state $s$ of SP $ P $ is said to be {\bf terminal} if there is no an edge 
outgoing from
$ at_P ^ s $.

Below we define a concept of a transition of a SP $ P $ corresponding to some action $ \alpha \in {\cal A} _P $.
This transition is a pair $( s, s') $ of  states from $ \Sigma_P $, the relationship between $s$ and $s'$ can be understood as follows: if $ P $  
is in the state $ s $ at the current time, then after sequential execution of EAs from $ \alpha $,  a state of $ P $ will be  $ s' $.

$\forall\,s, s'\in \Sigma_P $, $\forall\,\alpha\in {\cal A}_P$
the notation $s \ra {\alpha} s '$ denotes
the statement that $( s, s') $ is a transition corresponding to  $ \alpha$. This statement holds if
$ \alpha$ is a sequence of EAs  of the form $ \alpha_1 \ldots \alpha_n $, and
$$ \exists \, s_1, \ldots, s_ {n-1} \in \Sigma_P:
s \ral {\alpha_1} s_1, \;
s_1 \ral {\alpha_2} s_2, \;
\ldots, \;
s_ {n-1} \ral {\alpha_n} s', $$
where the statement $ s \ra {\alpha} s' $ in the case when $ \alpha $ is an EA is defined separately for each  form of  $ \alpha $, after the formal definition of this statement for each specific form of $ \alpha $ we informally interpret the state change as a result of this transition:
\bi
\i [(a)] if $ \alpha = c! e $, then $ \theta ^ {s'} = \theta ^ {s} $, and
 $$ [c ^ s] ^ {s'} = ([c ^ s] ^ s,
{e ^ s}), \; \;
 \forall \, c '\in {\cal D} _ {\bf C} \setminus \{c ^ s \} \; \;
[c '] ^ {s'} = [c'] ^ s,
$$
in this case, $ P $ sends the value $ e ^ s $ to the channel $ c ^ s $, after which
\bi 
\i the content of the channel $ c ^ s $ has been increased by adding $ e ^ s $,

\i
the binding of variables from ${\cal X}_P$ did not change, and contents
of all channels, except for the $ c ^ s $ channel, also did not change,
\ei 
\i [(b)] if $ \alpha = c? e $, then
$ [c ^ s] ^ s \neq \emptyset $ and
$$ \by \, \exists \, \theta \in \Theta (\hat X_P):
{(e ^ {\theta}) ^ {s}} = head ([c ^ s] ^ s),
\theta ^ {s'} = \theta \theta ^ {s}, \\
\, [c ^ s] ^ {s'} = tail ([c ^ s] ^ s), \; \;
\forall \, c '\in {\cal D} _ {\bf C} \setminus \{c ^ s \} \; \;
[c '] ^ {s'} = [c'] ^ s \ey $$
 in this case, $ P $ takes the value $ head ([c ^ s] ^ s) $ from the channel $ c ^ s $ and changes the current binding so that the value of the term $ e $  coincides with the accepted value on a new binding, after which \bi
\i the content of the channel $ c ^ s $ becomes $ tail ([c ^ s] ^ s) $,
\i contents of the other channels have not changed,
 \ei
\i [(c)] if $ \alpha = (e: = e ') $, then
$
\left\{\by
\exists \, \theta \in \Theta ({\hat X_P}):
{(e ^ {\theta}) ^ s} =
{(e ') ^ s}, \theta ^ {s'} = \theta \theta ^ {s}, \\
\forall \, c \in {\cal D} _ {\bf C} \; \;
[c] ^ {s'} = [c] ^ s, \ey \right.$\\
in this case, $ P $ changes the current binding so that the value of the 
term
$ e $  on the new binding would be equal to the value of the $ e '$ term on the old binding, contents of the channels does not change,

\i [(d)] if $ \alpha = [\! [\varphi] \!] $, then
$ \varphi ^ s = 1 $,
$ \theta ^ {s'} = \theta ^ {s} $, and $\forall \, c \in {\cal D} _ {\bf C} \; \;
[c] ^ {s'} = [c] ^ s$, \\
in this case, the binding and contents of the channels are not changed.
\ei

{\bf An empty transition} of SP $ P $ is a pair $(s,s')$
of states from $\Sigma_P $, such that $ \theta ^ s = \theta ^ {s'} $.
The empty transition is denoted by $ s \to s' $.

If a pair $ (s, s ')$ is a transition of a SP, then 
we say that this is a transition from $s$ to $s'$,
$ s $ and $s'$ is called a {\bf start} and an {\bf end} of this transition, respectively.

Let $ P $ be a SP, and $v\in P $, $v\neq P^0$. If  sets of all edges of $ P $ ending in $ v $ and starting at $ v $ are 
$ \{v_i \ral {\alpha_i} v \mid i = 1, \ldots, n \}$ and
$\{v \ral {\alpha'_i} v'_i \mid i = 1, \ldots, n '\}
$
respectively, where $ v$ differs from all the nodes $ v_i $ and $ v '_ {i} $, and either all actions $ \alpha_1, \ldots, \alpha_n $ are internal, or all actions $ \alpha'_1, \ldots , \alpha '_ {n'} $ are internal, then a {\bf reduction} operation can be applied to $P$, which consists of transforming this graph by
\bi \i removing the node $ v $ and associated edges, and \i adding edges of the form $ v_i \ral {\alpha_i \alpha '_ {i'}} v '_ {i'} $, where $ i = 1, \ldots, n, i '= 1, \ldots, n' $, and $ \alpha_i \alpha '_ {i'} $ is a concatenation of sequences $ \alpha_i $ and $ \alpha '_ {i'} $.
\ei

A {\bf renaming} is an  injective function $ \eta: X \to X '$, where $ X, X' \subseteq {\cal X} $.
For each renaming $ \eta: X \to X '$, each $ e \in {\cal E} $ and each SP $ P $, the notations $ e ^ {\eta} $ and $ P ^ \eta $ denote a term or a SP respectively, obtained from $ e $ or $ P $ by replacing $ \forall \, x \in X $ of each occurrence of $ x $ by $ \eta (x) $.
If $ P $ is a SP, and $ \eta $ is a renaming of the form
$ \eta:
\hat X_P
\to
{\it {\cal X}} \setminus X_P
$, then we will consider SP $ P $ and $ P ^ \eta $ as equal.

\subsection{Distributed Processes }

In this section we
introduce a concept of a distributed process, which can be used for
formal 
representation of MPI programs. 

A
{\bf distributed process (DP)} is a family of SPs
${\cal P} = \{P_i \mid i \in I \}$,
where  components of $ \{\hat X_ {P_i} \mid i \in I \} $ are disjoint and do not intersect with $ X _ {\cal P} \eam \bigcup_ {i \in I} X_ {P_i} $ (if this condition does not met, then we replace each SP $ P_i $ by an equal to it, in the sense defined at the end of  section
\ref{sdfdsfgdsgsersdvzx}, so that this condition will be met). Below we  assume that this condition does  met, even if  private variables of  SPs $ P_ {i} $ and $ P_ {i '} $ from $ {\cal P} $, where $ i \neq i' $, have the same designations.

For each DP ${\cal P}$
$ {\cal X} _ {\cal P} $ denotes the set of all variables occurred in $ {\cal P} $.

A {\bf state} of DP $ {\cal P} $ is a pair
$s = (
\theta ^ s,
\{c ^ s \mid c \in {\cal D} _ {\bf C} \})$,
where $ \theta ^ s \in \Theta ({{\cal X} _ {\cal P}}) ,
c^s\in {\cal D}^*$.
A set of all states of a DP $ {\cal P} $ is denoted by $ \Sigma _ {\cal P} $.

$ \forall \, s \in \Sigma _ {\cal P} $, $ \forall \,e \in {\cal E} ({\cal X} _ {\cal P}) $ the value $ e ^ {\theta ^ s} $ is denoted by $ e ^ s $.

Let 
${\cal P}= \{P_i \mid i \in I \}$ be a DP.
$\forall\,s\in \Sigma _ {\cal P}$, 
$ \forall \, i \in I$
$ s_i \eam (
\theta_i ^ s,
\{c ^ s \mid c \in {\cal D} _ {\bf C} \}) \in \Sigma_ {P_i} $,
where $ \theta_i ^ s \in \Theta ({\cal X} _ {P_i}) $,
$ \forall \, x \in {\cal X} _ {P_i} \; \; x ^ {\theta_i ^ s} =
x ^ {\theta ^ s} $.

 A state $ s \in \Sigma _ {\cal P} $ is said to be 
 {\bf initial} (and is denoted by $ 0 _ {\cal P} $), if $ \forall \, i \in I \; \; s_i = 0_ {P_i} $, 
{\bf terminal}, if $ \forall \, i \in I \; s_i $ is terminal, and
{\bf deadlock} if it is nonterminal, and $ \forall \, i \in I $ there is no non-empty transition of $ P_i $ from $ s_i $.

Let  $ {\cal P} = \{P_i \mid i \in I \} $ be a DP.
{\bf A transition} in $ {\cal P} $ corresponding to an action $ \alpha \in 
{\cal A} _ {P_i}$ is a pair $(s,s')$
of states from $\Sigma _ {{\cal P}} $, such what
\be {sdfadsf3wrf34}
s_i \ra {\alpha} s'_i, \;
\forall \, i '\in I \setminus \{i \} \; \;
{s_ {i '}} \to {s' _ {i'}}.
\ee

Property \re {sdfadsf3wrf34} is denoted by  $ P_i ^ {v \to v '}: s \to s' $, where $ v =at_{P_i}^s$, $ v'=at_{P_i}^{s'}$.
If $( s, s ')$ is a transition of $ {\cal P} $, then we will denote it by
$ s \to s' $.

The relationship between  states $ s, s' \in \Sigma _ {{\cal P}} $ satisfying  \re {sdfadsf3wrf34} can be interpreted as follows: if $ {\cal P} $ is in the state $ s $ at the current time, and from that moment on, a SP $ P_i \in {\cal P}$ sequentially performed  EAs from $ \alpha $, and $\forall\,i'\in I\setminus \{i\}$
$P_{i'}$ did not perform any actions during all this time, then after completion the execution of EAs from $ \alpha $, the new state of  $ {\cal P} $ is $ s' $.

The set $ \Sigma _ {{\cal P}} $ can be considered as a graph in which there is an edge from $ s $ to $ s' $ labeled $ \alpha_ {P_i} $
if and only if \re {sdfadsf3wrf34} is true. 
 
An {\bf execution} of a DP $ {\cal P} $ is a sequence of states $ s_0, s_1, \ldots $, such that $ s_0 = 0 _ {\cal P} $, and each pair $ s_i, s_ {i + 1} $ of neighboring states in this sequence is a transition of $ {\cal P} $.
 
 A state $ s $ of a DP $ {\cal P} $ is said to be 
 {\bf reachable} if there is a path from $ 0 _ {\cal P} $ to $ s $. Below $ \Sigma _ {\cal P} $ denotes the set of reachable states of $ {\cal P} $. 

 \subsection{A method for constructing 
a distributed process which is  
 a model of a MPI program}\label{metod}
 
We will consider only such MPI pro\-g\-rams that contain
\bi
\i operators of assignment, conditional transition, loop,
 \i functions of sending and receiving messages, defined in section \ref {peredachasoob}, and \i service MPI  functions (\verb'MPI_Init', etc.) mentioned in the program from section \ref {sdfdsfg3wergtegr}.
 \ei
 
Let $\Pi$ be a MPI program, 
$ X_ \Pi \subseteq {\cal X}$ be a set of input variables of $\Pi$, and
$ \varphi_ \Pi \in {\cal B} $ be 
an initial condition  of $ \Pi $.
A {\bf model} of $\Pi$ is the DP
${\cal P} _ \Pi = \{P_i \mid i \geq 0\}$,
where $ \forall \, i \geq 0$ SP $ P_i $ is constructed as follows.
\bi
\i
 $ {\cal X} _ {P_i} $ consists of variables from $ X_ \Pi $, and 
$ i $-th copies of private variables  of $ \Pi $,
$ X_ {P_i} = X_ \Pi $,
$ \varphi_ {P_i} = \varphi_ \Pi $.
\i The graph $ P_i $ is constructed by \bi \i
replacing in $ \Pi $ second argument of the function  \verb'MPI_Comm_rank' (in the MPI program presented in section \ref {sdfdsfg3wergtegr} this is  variable  \verb'rank') by $ i $, \i deleting non-executable parts of the resulting program,
and \i transformation the resulting program into graph form, similarly to how the program in operator form is transformed to a flowchart 
(with the difference that in flowcharts  actions are associated with nodes, 
and in our model actions are associated with edges).
\ei

\ei

Message passing functions are represented in $ P_i $ by the following actions:
\bi
\i  function \re {sdfsfsdfsend} is represented by the action $ c_r! (e, i, l) $, where \bi \i $ e $ is a term whose value must be equal to the 
content of the memory segment sent by this function,
\i $ r $ and $ l $ are the corresponding arguments of function
\re {sdfsfsdfsend}  (receiver number and tag, respectively), \ei
\i function \re {sdfsfsdfreceive} is represented in $ P_i $ by the action $ c_i? (e, s, l) $, where
\bi
\i $ e $ is a term whose value after performing this action must be equal to the received message,
\i $ s $ and $ l $ are new variables, \ei
and if the last argument of function \re {sdfsfsdfreceive}  has the name $ q $, then the expressions in $ P_i $ of the form $ q $\verb'.MPI_SOURCE' and $ q $\verb'.MPI_TAG' are replaced with $ s $ and $ l $, respectively,
\i representation of function \re {sdfsfsdfbcast} depends on $ i $:
\bi
\i for $ i = 0 $ the function \re {sdfsfsdfbcast} is represented by the action $ \circ! e $, where $ e $ is a term whose value must be equal 
to the content of the memory segment sent by this function,
\i for $ i \neq 0 $ the function \re {sdfsfsdfbcast} is represented by the action $ \circ? e $, where $ e $ is a term whose value after performing this action must be equal to the received message.
\ei

\ei

The reduction operation described in section \ref{sdfdsfgdsgsersdvzx}
can be applied to the constructed graph $ P_i $. 

To facilitate an analysis of  DP $ {\cal P} _ \Pi $, one can add to 
 actions of this DP assignments of the form $ \iota: = e $, where
 $ \iota $ is a new variable (called an {\bf auxiliary} variable), and
 $ e \in {\cal E} ({\cal X} _ {{\cal P} _ \Pi} \sqcup {\cal I}) $, where $ {\cal I} $ is a set of auxiliary variables.
The assignments of the above form $ \iota: = e $ are not actually performed actions, they are intended only to express dependencies between  values of variables during an execution of the DP. 
SPs from SPs $P_i\in  {\cal P} _ \Pi $ by adding assignments of the above form $ \iota: = e $, where $ \iota $ is an auxiliary variable, will be called {\bf augmented } SPs. 

\section{Modeling and verification of the matrix multiplication MPI program}

In this section, we apply the above concepts to modeling and verification of 
the matrix multiplication MPI program $\Pi$ described in section \ref {rejgkljktdhgdslfk}.
 
 \subsection {A model of the
  matrix multiplication MPI program}
 \label {sdfsfgw3gtgrsd}
 
Define  DP $ {\cal P} _ \Pi = \{P_i \mid i \geq 0 \} $, which is a model of
$\Pi$.

We assume that the following variables belong to $X_\Pi$:
$ A $, $ B $ (matrix factors), $ N $ (number of rows in $ A $),  \verb'nprocs'.

$ \forall \, i \geq 0 $,
when constructing SP $ P_i $, the following simplifications are used:
\bi
\i the function for input of matrix factors \verb'input (a, b)' in line 15 is executed once at the initial moment of execution of the root process, therefore in  SP  $ P_0 $ it is possible to omit  actions corresponding to this function, assuming that $ X_ {P_0} $ has  variables $ A $ and $ B $,  values of which are equal to  matrix factors,

\i BMP functions in $ P_0 $ in lines 16-17 and in $ P_i \; (\forall \, i \geq 1) $ in lines 44-45 are executed once, at the initial moment of execution of these SPs, so you can replace the corresponding actions of the form $ \circ! e $ and $ \circ? e $ on the assumption that $ B $ belongs to $ X_ {P_i} \; (\forall \, i \geq 1) $.
\ei

To shorten the notation in  SPs below, instead of  input variable \verb'nprocs', we  use  input variable $ n $, a value of which is equal to $ \verb'nprocs'-1 $.

 To build  SP $ P_0 $, only a part of $\Pi$ is used, located in lines 12-39.
 
 In  SP $ P_0 $ $ A $ and $ C $ are  names of arrays,  components of which are rows of corresponding matrices and are indexed by numbers $ 1, \ldots, N $, 
$\forall\,i=1,\ldots, N$  $ A_i $ and $ C_i $ denote
$ i $-th components of these arrays.

  $ P_0 $ has the following variables:
$ \by X_ {P_0} = \{A, B, N, n \}, \;
 \hat X_ {P_0} = \{C, i, j, k, l, p \}. \ey $

Initial condition:
$ \by \varphi_ {P_0} = & (N \geq 1) \wedge
(i = 1) \wedge (k = 0) \wedge (l = 1).
\ey $

 SP $ P_0 $ has the form 
 \be{z0}
\by
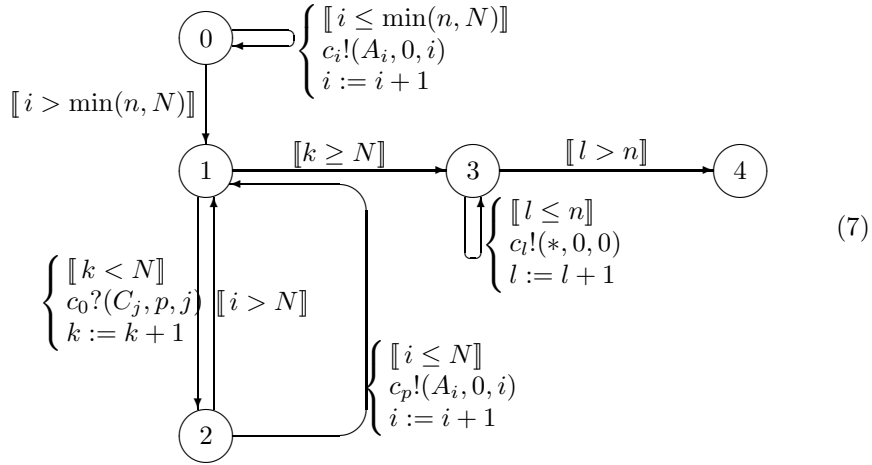
\begin{picture}(150,160)

\put(0,0){\oval(20,20)}
\put(0,0){\makebox(0,0)[c]{2}}
\put(0,100){\oval(20,20)}
\put(0,100){\makebox(0,0)[c]{1}}
\put(0,150){\oval(20,20)}
\put(0,150){\makebox(0,0)[c]{0}}
\put(100,100){\oval(20,20)}
\put(100,100){\makebox(0,0)[c]{3}}
\put(200,100){\oval(20,20)}
\put(200,100){\makebox(0,0)[c]{4}}

\put(0,140){\vector(0,-1){30}}
\put(-3,90){\vector(0,-1){80}}
\put(3,10){\vector(0,1){80}}
\put(10,100){\vector(1,0){80}}
\put(110,100){\vector(1,0){80}}

\put(10,153){\line(1,0){20}}
\put(30,147){\vector(-1,0){20}}
\put(30,150){\oval(6,6)[r]}

\put(97,90){\line(0,-1){20}}
\put(103,70){\vector(0,1){20}}
\put(100,70){\oval(6,6)[b]}

\put(150,102){\makebox(0,0)[b]{
$[\![\,l>n]\!]$
}}

\put(0,125){\makebox(0,0)[r]{
$[\![\,i>\min(n,N)]\!]$
}}

\put(50,100){\makebox(0,0)[b]{
$\begin{array}{lllll}
[\![k\geq N]\!]
\ey$
}}

\put(30,145){\makebox(0,0)[l]{
$\left\{\begin{array}{lllll}
[\![\,i\leq \min(n,N)]\!]\\
c_i!( A_{i},0,i)\\
i :=i+1
\ey\right.$
}}

\put(100,72){\makebox(0,0)[l]{
$\left\{\begin{array}{lllll}
[\![\,l\leq n]\!]\\
c_l!(*,0,0)\\l:=l+1
\ey\right.$
}}

\put(5,50){\makebox(0,0)[r]{
$\left\{\begin{array}{lllll}[\![\,k< N]\!]\\
c_0?( C_{ j},  p, j)\\
k:=k+1
\ey\right.$
}}

\put(0,50){\makebox(0,0)[l]{
$[\![\,i> N]\!]
$
}}

\put(10,0){\line(1,0){40}}
\put(60,10){\line(0,1){75}}
\put(50,95){\vector(-1,0){41}}
\put(50,10){\oval(20,20)[br]}
\put(50,85){\oval(20,20)[tr]}

\put(55,18){\makebox(0,0)[l]{
$\left\{\begin{array}{lllll}[\![\,i\leq N]\!]\\
c_p!(A_{i},0, i)
\\i :=i+1
\ey\right.$
}}

\end{picture}
\ey
\ee

In SP \re {z0}
\bi \i edge $ P_0 ^ {0 \to 0} $ corresponds to the loop in lines 18-22 of $\Pi$, the variable $ i $ in the label of this edge corresponds to the expression $ \verb'count'+ 1 $ in $\Pi$,
 \i the message sent by the \verb'MPI_Send 'function in lines
 21-22 of $\Pi$, is represented by the triple $ (A_i, 0, i) $,  third component of which is a tag of this message (i.e. a number of the corresponding row in the matrix $ A $),
\i edge $ P_0 ^ {0 \to 1} $ corresponds to the exit from this cycle, 
\i edges $ P_0 ^ {1 \to 2} $ and $ P_0 ^ {2 \to 1} $ correspond to a loop in lines 23-34 of $\Pi$,
 \i variable  $ k $ corresponds to  variable \verb'i' in this loop,
 \i the function \verb'MPI_Recv 'in lines 24-26 and the loop in lines 27-28 of $\Pi$ are replaced with a single action: receiving a message and writing it to the corresponding row of  matrix $ C $, \i edge $ P_0 ^ {1 \to 3 } $ corresponds to the exit from this loop,
 \i edges $ P_0 ^ {3 \to 3} $ and $ P_0 ^ {3 \to 4} $ correspond to the cycle in lines 35-37,
 \i  symbol $ * $ in the label of edge  $ P_0 ^ {3 \to 3} $ represents the empty string. \ei
 
To construct  SP $ P_i \;( i \geq 1) $, only the part of $\Pi$  located in lines 42-55 is used.
  $ P_i $ has the following variables: $ X_ {P_i} = \{B \}, \hat X_ {P_i} = \{Y_i, j_i \} $, where a value of $ B $ is the second factor matrix, and  values of $ Y_i $ are strings of real numbers. $ P_i $ has the following form: 
 \be{z1}
\by
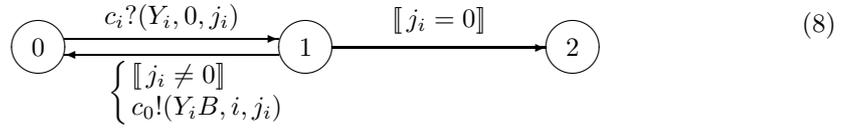
\begin{picture}(50,20)

\put(-100,0){\oval(20,20)}
\put(-100,0){\makebox(0,0)[c]{0}}
\put(0,0){\oval(20,20)}
\put(0,0){\makebox(0,0)[c]{1}}
\put(100,0){\oval(20,20)}
\put(100,0){\makebox(0,0)[c]{2}}

\put(-90,3){\vector(1,0){80}}
\put(-10,-3){\vector(-1,0){80}}
\put(10,0){\vector(1,0){80}}

\put(-50,5){\makebox(0,0)[b]{
$\by c_i?( Y_i,0, j_i)\ey$
}}
\put(50,5){\makebox(0,0)[b]{
$[\![\,j_i=0]\!]$
}}

\put(-40,-5){\makebox(0,0)[t]{
$\left\{\begin{array}{llll} [\![\,j_i\neq 0]\!]\\c_0!(Y_i B,i,j_i)\\
\ey\right.$
}}

\end{picture}
\ey\vspace{5mm}
\ee

\subsection{Verification of a distributed process that 
is a model of 
the  matrix multiplication MPI program}

To prove the statement that DP $ {{\cal P} _ \Pi} $ defined in section \ref {sdfsfgw3gtgrsd}, satisfies  specification \re {sdfvdsgg3r435grasdv}, we  introduce auxiliary variables $ \alpha $, $ \beta $, $ \gamma $,
and actions related  to these variables 
that do not affect an execution of the DP.
Values of the auxiliary variables  have the following meaning:
their initial values are $ \emptyset $, and 
$ \forall \, s \in \Sigma _ {{\cal P} _ \Pi} $
\bi

\i $ \alpha ^ s \subseteq \{1, \ldots, n \} $, $ \alpha ^ s $ consists of channel numbers from $ \{c_1, \ldots, c_n \} $ with non-empty content in state $ s $,
\i $ \beta ^ s \subseteq \{1, \ldots, N \} $, $ \beta ^ s $ consists of  numbers of  rows of $ A $, for which their product by $ B $ is calculated in state $ s $,
\i
$ \gamma ^ s \subseteq \{1, \ldots, N \} $, $ \gamma ^ s $ is the set of numbers of  rows that $ P_0 $ wrote in $ C $ 
during an execution of $ {{\cal P } _ \Pi} $ up to state $ s $.

\ei

Augmented SP $ P_0 $ has the form 
 \be{z0aug}
\by
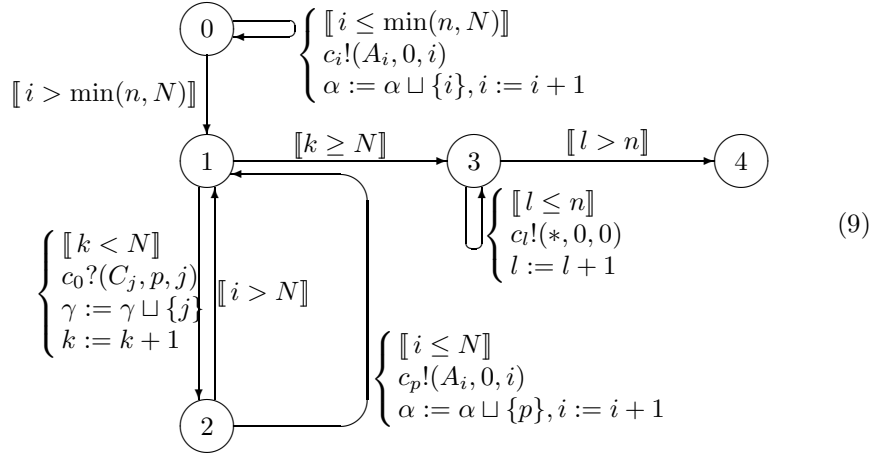
\begin{picture}(150,155)

\put(0,0){\oval(20,20)}
\put(0,0){\makebox(0,0)[c]{2}}
\put(0,100){\oval(20,20)}
\put(0,100){\makebox(0,0)[c]{1}}
\put(0,150){\oval(20,20)}
\put(0,150){\makebox(0,0)[c]{0}}
\put(100,100){\oval(20,20)}
\put(100,100){\makebox(0,0)[c]{3}}
\put(200,100){\oval(20,20)}
\put(200,100){\makebox(0,0)[c]{4}}

\put(0,140){\vector(0,-1){30}}
\put(-3,90){\vector(0,-1){80}}
\put(3,10){\vector(0,1){80}}
\put(10,100){\vector(1,0){80}}
\put(110,100){\vector(1,0){80}}

\put(10,153){\line(1,0){20}}
\put(30,147){\vector(-1,0){20}}
\put(30,150){\oval(6,6)[r]}

\put(97,90){\line(0,-1){20}}
\put(103,70){\vector(0,1){20}}
\put(100,70){\oval(6,6)[b]}

\put(150,102){\makebox(0,0)[b]{
$[\![\,l>n]\!]$
}}

\put(0,125){\makebox(0,0)[r]{
$[\![\,i>\min(n,N)]\!]$
}}

\put(50,100){\makebox(0,0)[b]{
$\begin{array}{lllll}
[\![k\geq N]\!]
\ey$
}}

\put(30,140){\makebox(0,0)[l]{
$\left\{\begin{array}{lllll}
[\![\,i\leq \min(n,N)]\!]\\
c_i!( A_{i},0,i)\\
\alpha:=\alpha\sqcup\{i\},
i :=i+1
\ey\right.$
}}

\put(100,72){\makebox(0,0)[l]{
$\left\{\begin{array}{lllll}
[\![\,l\leq n]\!]\\
c_l!(*,0,0)\\l:=l+1
\ey\right.$
}}

\put(5,50){\makebox(0,0)[r]{
$\left\{\begin{array}{lllll}[\![\,k< N]\!]\\
c_0?( C_{ j},  p, j)\\
\gamma:=\gamma\sqcup\{j\}\\k:=k+1
\ey\right.$
}}

\put(0,50){\makebox(0,0)[l]{
$[\![\,i> N]\!]
$
}}

\put(10,0){\line(1,0){40}}
\put(60,10){\line(0,1){75}}
\put(50,95){\vector(-1,0){41}}
\put(50,10){\oval(20,20)[br]}
\put(50,85){\oval(20,20)[tr]}

\put(58,18){\makebox(0,0)[l]{
$\left\{\begin{array}{lllll}[\![\,i\leq N]\!]\\
c_p!(A_{i},0, i)
\\
\alpha:=\alpha\sqcup\{p\}, 
i :=i+1
\ey\right.$
}}

\end{picture}
\ey\vspace{2mm}
\ee

$\forall\,i=1,\ldots, n$
 augmented SP $ P_i $ has the form 
 \be{z1aug}
\by
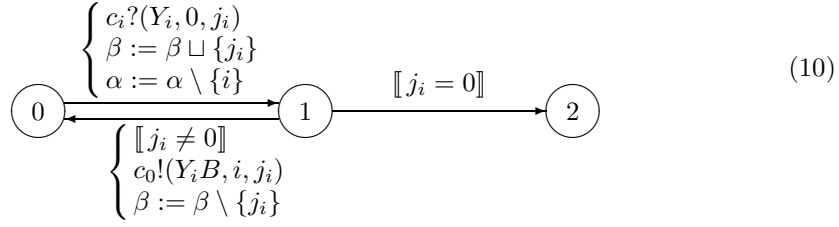
\begin{picture}(50,35)

\put(-100,0){\oval(20,20)}
\put(-100,0){\makebox(0,0)[c]{0}}
\put(0,0){\oval(20,20)}
\put(0,0){\makebox(0,0)[c]{1}}
\put(100,0){\oval(20,20)}
\put(100,0){\makebox(0,0)[c]{2}}

\put(-90,3){\vector(1,0){80}}
\put(-10,-3){\vector(-1,0){80}}
\put(10,0){\vector(1,0){80}}

\put(-50,5){\makebox(0,0)[b]{
$\left\{\by c_i?( Y_i,0, j_i)\\
\beta:=\beta\sqcup\{j_i\}\\
\alpha:=\alpha\setminus\{i\}\ey\right.$
}}
\put(50,5){\makebox(0,0)[b]{
$[\![\,j_i=0]\!]$
}}

\put(-40,-5){\makebox(0,0)[t]{
$\left\{\begin{array}{llll} [\![\,j_i\neq 0]\!]\\c_0!(Y_i B,i,j_i)\\
\beta:=\beta\setminus\{j_i\}
\ey\right.$
}}

\end{picture}
\ey
\ee\\$\;$\\$\;$

The use of $ \sqcup $
in \re {z0aug} and \re {z1aug}  for set union operations expresses the statement (following from the following theorem \ref {d2eq}) that whenever these operations are performed, their arguments are actually disjoint sets.\\

\refstepcounter {theorem}
{\bf Theorem \arabic{theorem}\label{d2eq}}.

$ \forall \, s \in \Sigma _ {{\cal P} _ \Pi} $, if $ at ^ s_ {P_0} \neq 3,4 $, then the following statements are true: 
\bn
\i \label {utver1}
$
\alpha ^ s \subseteq \{1, \ldots, i ^ s-1 \} $,
\i \label {utver2}
$ i ^ s-1 \leq N $
\i \label {utver3}
$ | \gamma ^ s | = k ^ s \leq N $,
\i \label {utver4} if $ at_ {P_0} ^ s = 1 $ and
$ k ^ s <N $, then $ k ^ s <i ^ s-1 $,
\i \label {utver5}
$ [c_0] ^ s_2 \cap \alpha ^ s = \emptyset $,
\i \label {utver6}
$ \forall \, i = 1, \ldots, n $ \bn \i \label {dsfsgdasgas}
$ | [c_i] ^ s | = 1$, if $ i \in \alpha ^ s $, and
$ | [c_i] ^ s | = 0$, otherwise,
\i $ at_ {P_i} ^ s = 1 \; \Rightarrow \;
(i \not \in \alpha ^ s) \wedge (j ^ s_i \not \in \gamma ^ s) $
\en
\i \label {utver7}
$ at_ {P_0} ^ s = 2 \; \Rightarrow \; p ^ s \not \in \alpha ^ s $,
$ p ^ s \in \{1, \ldots, i ^ s-1 \} $,
\i \label {utver8}
$ [c_1] _3 ^ s \sqcup \ldots \sqcup [c_n] _3 ^ s
\sqcup \beta ^ s \sqcup [c_0] _3 ^ s
\sqcup \gamma ^ s
= \{1, \ldots, i ^ s-1 \}
$,
\i \label {utver9}
$ \forall \, p
\in \alpha ^ s \; \; [c_p] ^ s $ has the form
$ \{(A_i, 0, i) \} $, where $ i \in \{1, \ldots, N \} $,
\i \label {utver19}
each element of $ [c_0] ^ s $ has the form $ (A_iB, p, i) $, where $ i \in \{1, \ldots, N \} $,
\i \label {utver10} $ \forall \, j \in \gamma ^ s \; \;
C_j = A_jB $,
\i \label {utver11} $ k ^ s = N \; \Rightarrow \; \gamma ^ s = \{1, \ldots, N \} $.
\en

{\bf Proof}.

All the  statements are substantiated inductively: they 
are  true in $ 0 _ {{\cal P} _ \Pi} $, and 
retain their truth after each transition $ s \to s' $ of $ {{\cal P} _ \Pi} $, where $ at_ {P_0} ^ {s'} \neq 3,4 $.
$ \blackbox $ \\

\refstepcounter {theorem}
{\bf Theorem \arabic {theorem}\label{d2e2132q}.}

There are no deadlocks in $ \Sigma _ {{\cal P} _ \Pi} $. \\

{\bf Proof}.

Let there is a deadlock $ s \in 
 \Sigma _ {{\cal P} _ \Pi} $. It is not hard to prove that $ at_ {P_0} ^ s 
 \not\in \{0, 2, 3\}$, and $ \forall \, i = 1, \ldots, n $ $ at_ {P_i} ^ s \neq 1$.
So, $ at_ {P_0} ^ s \in \{ 1, 4\}$.
\bn \i
Let $ at_ {P_0} ^ s = 1 $. Since $s$ is a deadlock, then
$ k ^ s <N $, whence  $ [c_0] ^ s = \emptyset $.

From  statement \ref {utver4} of theorem \ref {d2eq} it follows that $ k ^ s <i ^ s-1 $, whence, based on statement \ref {utver8} of theorem \ref {d2eq} and the equality $ [c_0] ^ s = \emptyset $ we get:
\be {dsfgdsgdsfds}
[c_1] _3 ^ s \sqcup \ldots \sqcup [c_n] _3 ^ s
\sqcup \beta ^ s \neq \emptyset.
\ee

If $ \exists \, i \in \{1, \ldots, n \} $: $ [c_i] ^ s \neq \emptyset $, then the 
assumption that 
$ s $ is a deadlock implies that $ at_ {P_i} ^ s \neq 0 $, therefore $ at_ {P_i} ^ s = 2 $, whence it is easy to get that $ at_ {P_0} ^ s = 4 $.
This is possible only if $ k ^ s \geq N $, which contradicts the  inequality $ k ^ s <N $.

Therefore, $ \forall \, i \in \{1, \ldots, n \} \; [c_i] ^ s = \emptyset $, and from \re {dsfgdsgdsfds} it follows that $ \beta \neq \emptyset $.
By analyzing  SP $ P_i \; (i = 0, \ldots, n) $ it is easy to prove that this is possible only if $ \exists \, i \in \{1, \ldots, n \} $: $ at_ {P_i } ^ s = 2 $. As stated above, this  is impossible.

\i Let $ at_ {P_0} ^ s = 4 $.
Then $ k ^ s \geq N $, whence $ | \gamma ^ s | = k ^ s = N $.
Let $ s '$ be  first state on a path $ \pi $ from $ 0 _ {{\cal P} _ \Pi} $ to $ s $, such that $ k ^ {s'} = N $.
It is easy to see that $ at_ {P_0} ^ {s'} = 2 $.
From  statements \ref {utver2} and \ref {utver8} of  theorem \ref {d2eq}
it follows that $ i ^ {s'} - 1 = N $, and
$$[c_0] ^ {s'} =
[c_1] ^ {s '} = \ldots = [c_n] ^ {s'} =
\beta ^ {s'} = \emptyset,\;\;
at_ {P_1} ^ {s'} = 0, \ldots,
at_ {P_n} ^ {s'} = 0.$$

The only transition from $ s' $ 
corresponds to the action 
$ [\! [i> N] \!] $,  
and this transition has the form 
$ P_0 ^ {2 \to 1}: s' \to s''$.
The only transition from $ s''$ 
corresponds to the action $ [\! [k \geq N] \!] $ 
and this transition has the form $ P_0 ^ {1 \to 3}: s''\to s''' $.
It is easy to see that all states 
in the tail of  $ \pi $ starting from $ s''' $ 
are not deadlocks.
\en 

In both cases, we get a contradiction, which is a consequence of the assumption that $ \Sigma _ {{\cal P} _ \Pi} $ has a deadlock state.
Thus, there are no deadlock states in $ \Sigma _ {{\cal P} _ \Pi} $.
$ \blackbox $ \\

\refstepcounter {theorem}
{\bf Theorem \arabic{theorem}\label{d2e24561q}.}

Any execution of  $ {\cal P}_\Pi $  terminates after a finite sequence of steps. \\

{\bf Proof}.

Let there is an infinite execution $ \pi $ of  $ {\cal P}_\Pi $.
Prove that a number of transitions in $ \pi $ corresponding to  actions of $ P_0 $ is finite.
If this is not so, then \bi \i there are no states $ s \in \pi$ such that $ at_ {P_0} ^ s = 3 $, \i there is a state $ s_1 \in \pi $, such that $ at_ {P_0} ^ {s_1} = 1 $,
\i each transition in  $ \pi $, starting from $ s_1 $, corresponding to some action $ P_0 $, either  has the form $ P_0 ^ {1 \to 2}: s \to s' $, or 
has the form  $ P_0 ^ { 2 \to 1}: s \to s '$, and the number of transitions of the form $ P_0 ^ {1 \to 2}: s \to s' $ is infinite. This is impossible due to the fact that each such transition increases the value of the variable $ k $, which, according to  statement \ref {utver3} of  theorem \ref {d2eq}, is bounded  by $ N $.
\ei

Let $ s '\in \pi $ be a state starting from which $ \pi $ does not contain transitions corresponding to actions of $ P_0 $, and $ \pi' $ is a tail of $ \pi $,
starting with $ s' $. It is not hard to see that
\be {sdfgadsgew45}
\by
\mbox {$ \exists \, i \in \{1, \ldots, n \} $:
$ \pi '$ contains infinitely many} \\\mbox {transitions of the form $ P_i ^ {0 \to 1}: s \to s' $.} \ey \ee
Because $ \pi '$ does not contain transitions corresponding to actions of $ P_0 $, then the value of $ | [c_i] ^ s | $ cannot increase, and according to \re {sdfgadsgew45} it decreases infinitely, which is impossible.
$ \blackbox $ \\

It follows from the above theorems that each execution of DP $ {\cal P} _ \Pi $ is finite and terminates in some terminal state $ s $.
From $ at_ {P_0} ^ s = 4 $ it follows that $ | \gamma ^ s | = k ^ s = N $, whence, according to  statements \ref {utver10} and \ref {utver11} of  theorem \ref {d2eq}, it follows that  DP $ {\cal P} _ {\Pi} $ satisfies the specification stated in section\ref {sdafadsfgaetga}. $ \blackbox $ 

\section{Conclusion}

In this article, we introduced a new mathematical model of parallel programs, 
proposed an approach for verifying such programs, and considered an example of verifying a matrix multiplication program based on the proposed model. 
The main advantage of this approach is the possibility of its application for  programs that generate a  unlimited set of SPs.

Problems for further research related to the proposed model can be the following:
1). to expand the proposed model by introducing concepts for modeling synchronous message passing, and other mechanisms for organizing parallel execution, 2).
to introduce specification language of properties of DPs, in which properties of parallel programs are expressed in terms of observational equivalence, and to 
elaborate algorithms for recognizing observational equivalence.

\end{document}